\documentclass[twocolumn,aps,prc,tightenlines,floats,floatfix,nofootinbib]{revtex4}%%

\def\bea{\begin{eqnarray}}
\def\eea{\end{eqnarray}}

\def\sfrac#1#2{{\textstyle \frac{#1}{#2}}}

\def\be{\begin{equation}}
\def\ee{\end{equation}}
\def\ba{\begin{eqnarray}}
\def\ea{\end{eqnarray}}

\def\st#1{{\kern-4pt} \not\!#1}

\usepackage{graphics}
\usepackage{graphicx}
\usepackage{epsf}
\usepackage{amsmath}
\usepackage{amssymb}
\begin{document}

\phantom{0}
\vspace{-0.2in}
\hspace{5.5in}
\parbox{1.5in}{ %}
\leftline{ADP-11-17/T739}}

\vspace{-1in}%\parbox{1.5in}{ \vspace{-9.6in}}  % moves the preprint box down

\title
{\bf A simple relation between the $\gamma N \to N(1535)$
helicity amplitudes}

\author{G. Ramalho$^{1}$ and K. Tsushima$^2$
\vspace{-0.1in}  }

\affiliation{
$^1$CFTP, Instituto Superior T\'ecnico,
Universidade T\'ecnica de Lisboa,
Av.~Rovisco Pais, 1049-001 Lisboa, Portugal
\vspace{-0.15in}}
\affiliation{$^2$CSSM, School of Chemistry and Physics,
University of Adelaide, Adelaide SA 5005, Australia}

\vspace{0.2in}
\date{\today}

\phantom{0}

\begin{abstract}
It is shown that the helicity amplitudes $A_{1/2}$ and $S_{1/2}$
in the $\gamma N \to N(1535)$ reaction,
 can be well related by
$S_{1/2} = -\frac{\sqrt{1+\tau}}{\sqrt{2}} \frac{M_S^2-M^2}{2 M_S Q} A_{1/2}$
in the region $Q^2 > 2$ GeV$^2$,
where $M$ and $M_S$ are the nucleon and $N(1535)$
masses, $q^2=-Q^2$ the four-momentum transfer squared,
and $\tau=\sfrac{Q^2}{(M_S+M)^2}$.
This follows from the fact that the Pauli-type transition form factor $F^*_2$
extracted from the experimental data,
turns up to show
$F_2^\ast \simeq 0$ for $Q^2 > 1.5$ GeV$^2$.  
The observed relation is tested by 
the experimentally extracted helicity amplitudes
and the MAID parametrization.
A direct consequence of the relation is that, the assumption
$|A_{1/2}| \gg |S_{1/2}|$ is not valid for high $Q^2$.
Instead, both amplitudes $A_{1/2}$ and $S_{1/2}$ have the same
$Q^2$ dependence in the high $Q^2$ region,
aside from that $S_{1/2}$ has an extra factor,
$- \sfrac{1}{\sqrt{2}}\sfrac{M_S-M}{2M_S}$.
The origin of this relation is interpreted
in a perspective of a quark model.
\end{abstract}

%\pacs{14.20.Gk, 13.40.Gp, 12.39.Ki}% PACS, the Physics and Astronomy
                                   % Classification Scheme.

%\phantom{0}
%\vspace{7.0in}
%\vspace{-6in}
\vspace*{0.9in}  % sets how far the title is below the preprint box
\maketitle

%\section{Introduction}

The electroproduction of a spin 1/2 baryon resonance $N^\ast$
on a nucleon ($\gamma^\ast N \to N^\ast$) is described by the two
independent helicity amplitudes, $A_{1/2}$ and $S_{1/2}$,
which depend on the initial and final state polarizations.
While the helicity is conserved in the transverse amplitude $A_{1/2}$,
it is changed by one unit in the longitudinal amplitude $S_{1/2}$.
These amplitudes are frame dependent.
For the transition between the nucleon state
$\left| N, S_z=\pm \sfrac{1}{2} \right>$ (mass $M$) and the spin-1/2 nucleon
resonance state $\left| N^\ast, S_z^\prime=\pm \sfrac{1}{2} \right>$
(mass $M_R$),
one can define the helicity amplitudes
in the $N^\ast$ rest frame
in terms of the transition current $J^\mu$ and
the photon polarization vector $\varepsilon^{(\lambda)}_\mu$
with $\lambda=0$ (longitudinal) or $\lambda=\pm 1$ (transverse)
\cite{Aznauryan08}:
\ba
& &
\hspace{-.5cm}
A_{1/2}= \sqrt{\frac{2\pi \alpha}{K}}
\left< N^\ast,S_z^\prime=+ \frac{1}{2} \right|
\varepsilon^{(+)} \cdot J
\left| N, S_z= - \frac{1}{2} \right>, \nonumber \\
%& & \label{eqA120}
& &
\hspace{-.5cm}
S_{1/2}= \sqrt{\frac{2\pi \alpha}{K}}
\left< N^\ast,S_z^\prime=+ \frac{1}{2} \right|
\varepsilon^{(0)} \cdot J
\left| N, S_z= + \frac{1}{2}  \right>
\frac{|{\bf q}|}{Q},
\nonumber \\
& &
\label{eqS120}
\ea
where $\alpha$ is the electromagnetic fine structure constant, and
\be
K= \frac{M_R^2-M^2}{2M_R}.
\ee
In the above $q^2=-Q^2$ is the four-momentum transfer squared and
${\bf q}$ the photon three-momentum in
the $N^\ast$ rest frame.
For simplicity we write the current $J^\mu$
in units of $e=\sqrt{4\pi \alpha}$.
%%%of magnitude of the electron charge $e=\sqrt{4\pi \alpha}$.

The helicity amplitudes
in the $N^\ast$ rest frame (\ref{eqS120})
are defined by the transition current
$J^\mu$, which can be defined in terms
of the two independent Dirac-type ($F_1^\ast$)
and Pauli-type ($F_2^\ast$) form factors,
which are frame independent (covariant) and
exclusive functions of $Q^2$
[the superscript ($^*$) is introduced to indicate
the final state is a nucleon excited state $N^\ast$].

Although the analysis of the nucleon system is
usually made in terms of the covariant electromagnetic form factors
\cite{Nucleon,Perdrisat07},
the data related with nucleon resonances are
mostly represented in terms of the helicity amplitudes
\cite{MAID,CLAS,Dalton09}
with the exception for the reaction
$\gamma N \to N(1440)$~\cite{Aznauryan08,Roper}.

%%%%%%%%%%%%%%%%%%%%%%%%%%%%%%%%%%%%%%%%%%%%%%%%%%%%%%%%%%%%%
% SECTION
%%%%%%%%%%%%%%%%%%%%%%%%%%%%%%%%%%%%%%%%%%%%%%%%%%%%%%%%%%%%%

%\section{$\gamma N \to N(1535)$ form factors and helicity amplitudes}

In this work we study the
$\gamma N \to N(1535)$ reaction
based on the covariant form factor representation.
The empirically extracted results for the transition form factors
will lead to a new, simple, and important relation
between the helicity amplitudes.
In the literature the amplitude $S_{1/2}$ is
generally neglected compared to $A_{1/2}$.
It was only recent that both the $S_{1/2}$ and $A_{1/2}$
amplitudes were extracted
simultaneously in the analysis of the cross section data.
Presently, we have results for the both amplitudes
from a MAID analysis of old data~\cite{MAID},
and recent results from CLAS (Jefferson Lab)~\cite{CLAS}.

The $N(1535)$ resonance is an $S_{11}$ state with
negative parity.
The transition current for $\gamma N \to S_{11}$   
can be represented~\cite{S11} as
\be
J^\mu=\bar u_S(P_+)
\left[
F_1^\ast\left(
 \gamma^\mu - \frac{\st q q^\mu}{q^2} \right) + F_2^\ast
\frac{i \sigma^{\mu \nu} q_\mu}{M_S+ M}
\right]\gamma_5 u(P_-),
\label{eqJS}
\ee
where the subscript $S$ stands for the quantities
associated with $S_{11}$.
We note that there are several equivalent definitions for the
transition current~\cite{Aznauryan08,Braun09,Pace99},
but we choose the present one, due to the
similarity with the nucleon and $N(1440)$ cases.

\begin{figure}[t]
%\vspace{.4cm}
\centerline{
\mbox{
\includegraphics[width=3.0in]{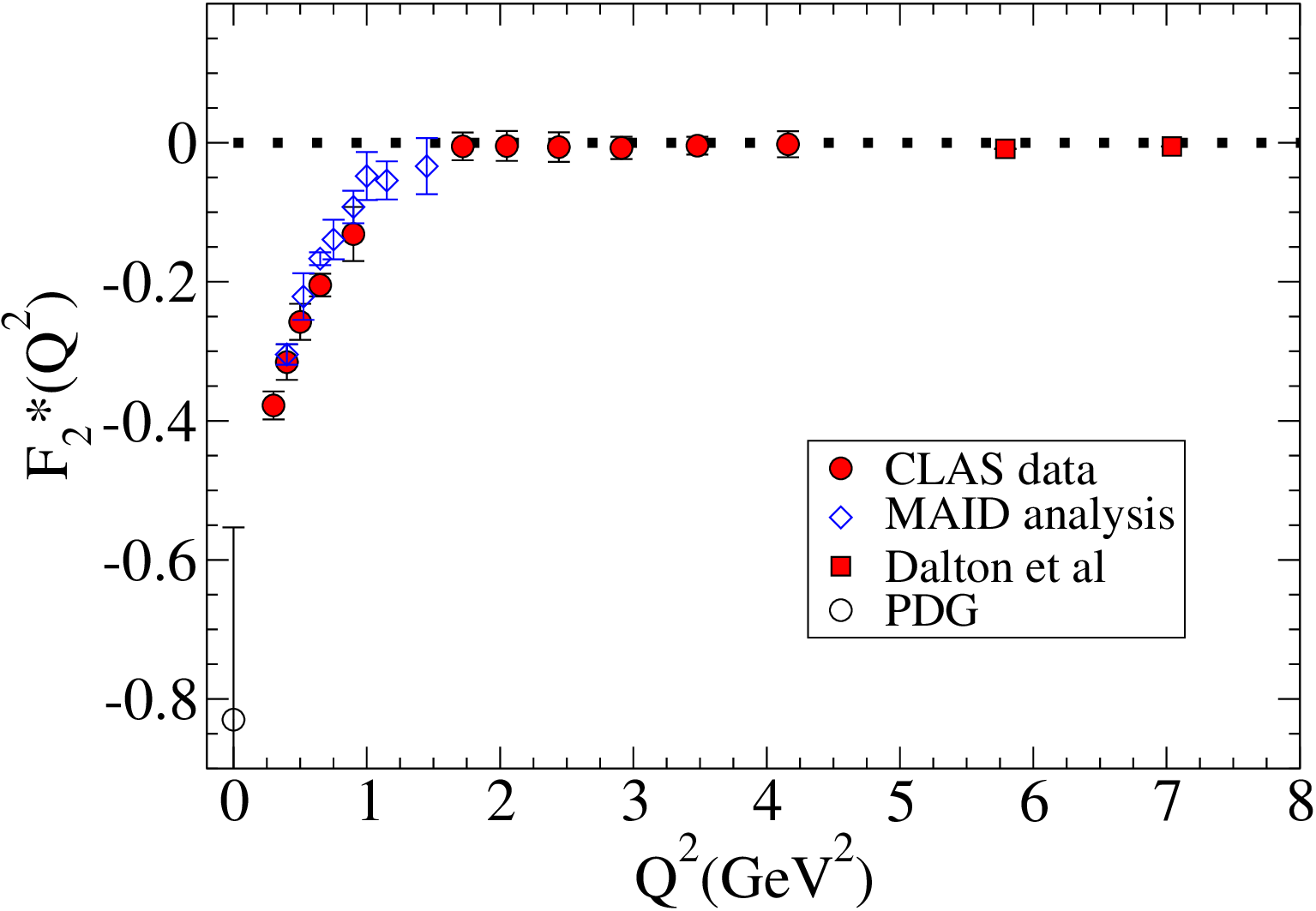} } }
\vspace{.3cm}
\centerline{
\mbox{
\includegraphics[width=3.0in]{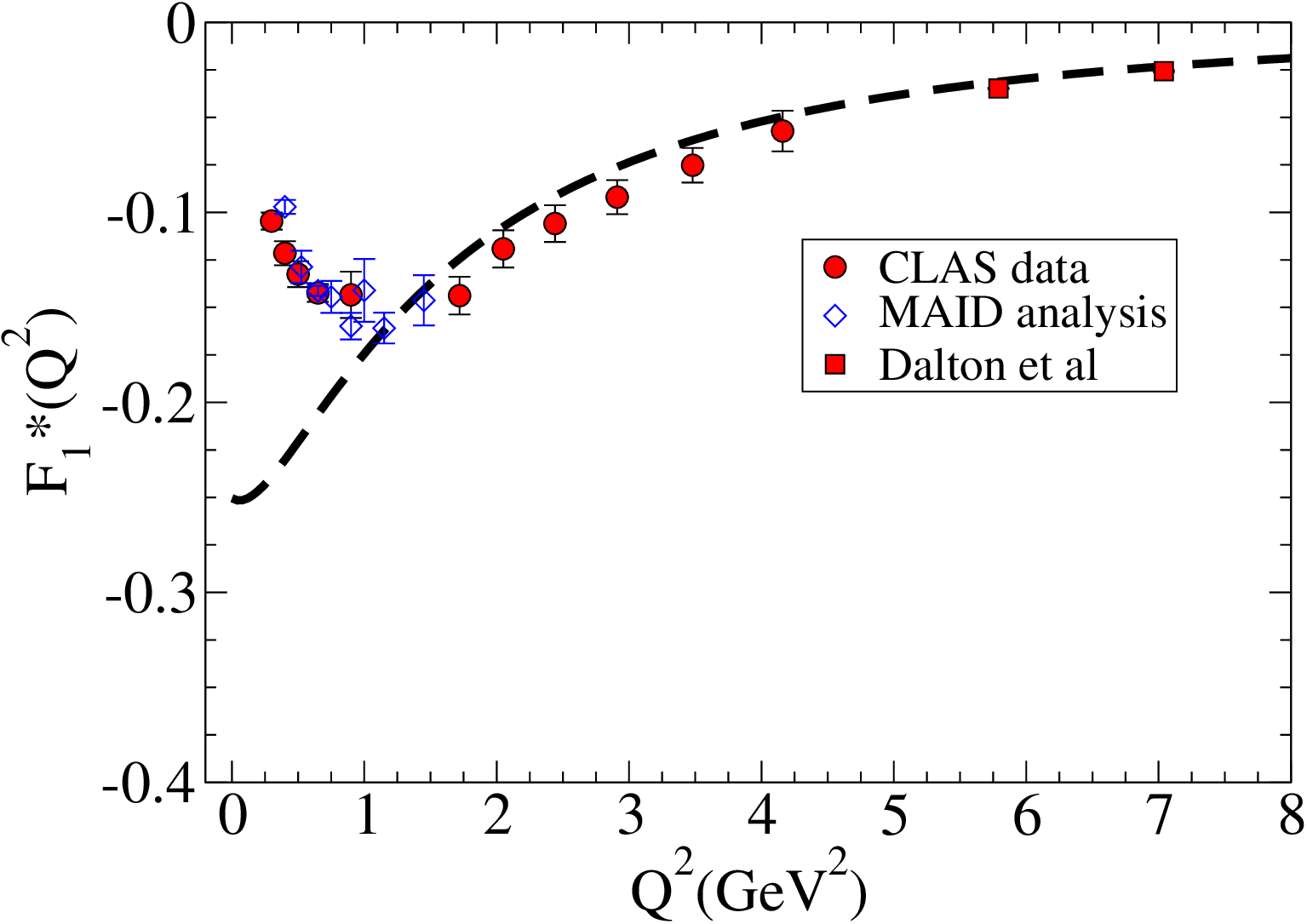} } }
\caption{\footnotesize
$\gamma N \to N(1535)$ transition form factors.
The dashed-line for $F_1^\ast$ represents the spectator quark
model result from Ref.~\cite{S11}.
Data are from Refs.~\cite{CLAS,MAID,Dalton09}.}
\label{figS11}
\end{figure}

The transition form factors, $F_1^\ast$ and $F_2^\ast$
extracted from the data
for the $\gamma N \to N(1535)$ reaction,
are shown in Fig.~\ref{figS11}.
A remarkable observation in the figure
is that, $F_2^\ast \simeq  0$ for $Q^2 > 1.5$ GeV$^2$.
Although this is due to the definition
of $F_2^\ast$ in the current Eq.~(\ref{eqJS}),
it has an important consequence when expressed
in terms of the helicity amplitudes
$A_{1/2}$ and $S_{1/2}$ as we will show next.

The relations for the form factors and
the helicity amplitudes are given by,
\ba
F_1^\ast&=&
-\frac{Q^2}{Q_-^2}
\frac{1}{2b} \left[
A_{1/2} - \sqrt{2} \frac{M_S-M}{|{\bf q}|}S_{1/2}
\right], \label{eqF1S}  \\
F_2^\ast &=&
-\frac{M_S^2-M^2}{Q_-^2}   \nonumber \\
& &
\times
\frac{1}{2b} \left[
A_{1/2} + \sqrt{2} \frac{Q^2}{(M_S-M)|{\bf q}|}S_{1/2}
\right],
\label{eqF2S}
\ea
where $Q_\pm^2=(M_S\pm M)^2+Q^2$, and
\be
b= e \sqrt{\frac{Q_+^2}{8M (M_S^2-M^2)}}.
\ee
In the above
\be
|{\bf q}|= \frac{\sqrt{Q_+^2Q_-^2}}{2M_S},
\label{eqq2}
\ee
is the absolute value of the photon three-momentum
in the $S_{11}$ rest frame.

From Eq.~(\ref{eqF2S}), 
the condition $F_2^\ast \simeq 0$ is equivalent to 
\be
S_{1/2} \simeq - \frac{1}{\sqrt{2}} \frac{(M_S-M) |{\bf q}|}{Q^2} A_{1/2}.
\label{eqS12a}
\ee
We can simplify Eq.~(\ref{eqq2})
for $Q^2 \gg $ $(M_S-M)^2 \simeq 0.355$ GeV$^2$ as
\be
|{\bf q}| \simeq \sqrt{1+ \tau}
 \frac{(M_S+M)}{2 M_S}  Q,
\label{eqQap}
\ee
where $\tau= \sfrac{Q^2}{(M_S+M)^2}$.
This approximation has a precision better than
10\% for $Q^2 > 1.8$ GeV$^2$.
Combining Eqs.~(\ref{eqS12a}) and~(\ref{eqQap}),
we obtain a simple relation,
\be
S_{1/2} \simeq -\frac{\sqrt{1+\tau}}{\sqrt{2}} \frac{M_S^2-M^2}{2 M_S Q}
A_{1/2},
\label{eqScaling}
\ee
for the region $Q^2 > 1.8$ GeV$^2$.
The relation of Eq.~(\ref{eqScaling}) is the
main result of this work.
We call this relation by {\it scaling}
(between $S_{1/2}$ and $A_{1/2}$).

Another interesting point
concerning the $\gamma N \to N(1535)$ reaction
is the pQCD estimate of $Q^3 A_{1/2}$
in the high $Q^2$ region~\cite{Carlson98},
where it gives a magnitude much larger
than the other resonance cases, e.g.,
than for the $\gamma N \to \Delta$ reaction~\cite{Carlson98}.
However, we  will not discuss this point here.
(See Refs.~\cite{S11,Dalton09} for more details).

\begin{figure}[t]
%\vspace{.4cm}
\centerline{
\mbox{
\includegraphics[width=3.0in]{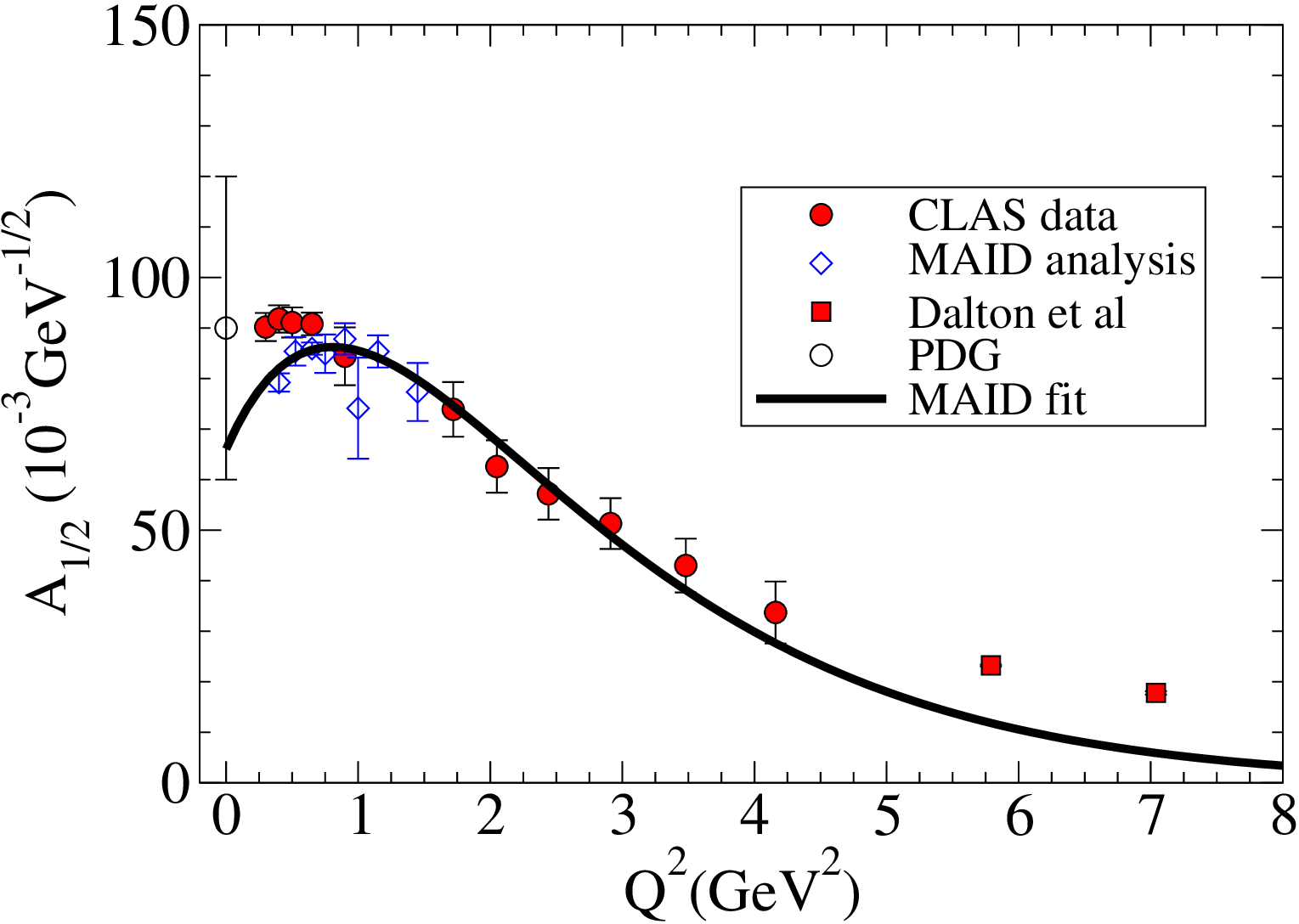} } }
\vspace{.3cm}
\centerline{
\mbox{
\includegraphics[width=3.0in]{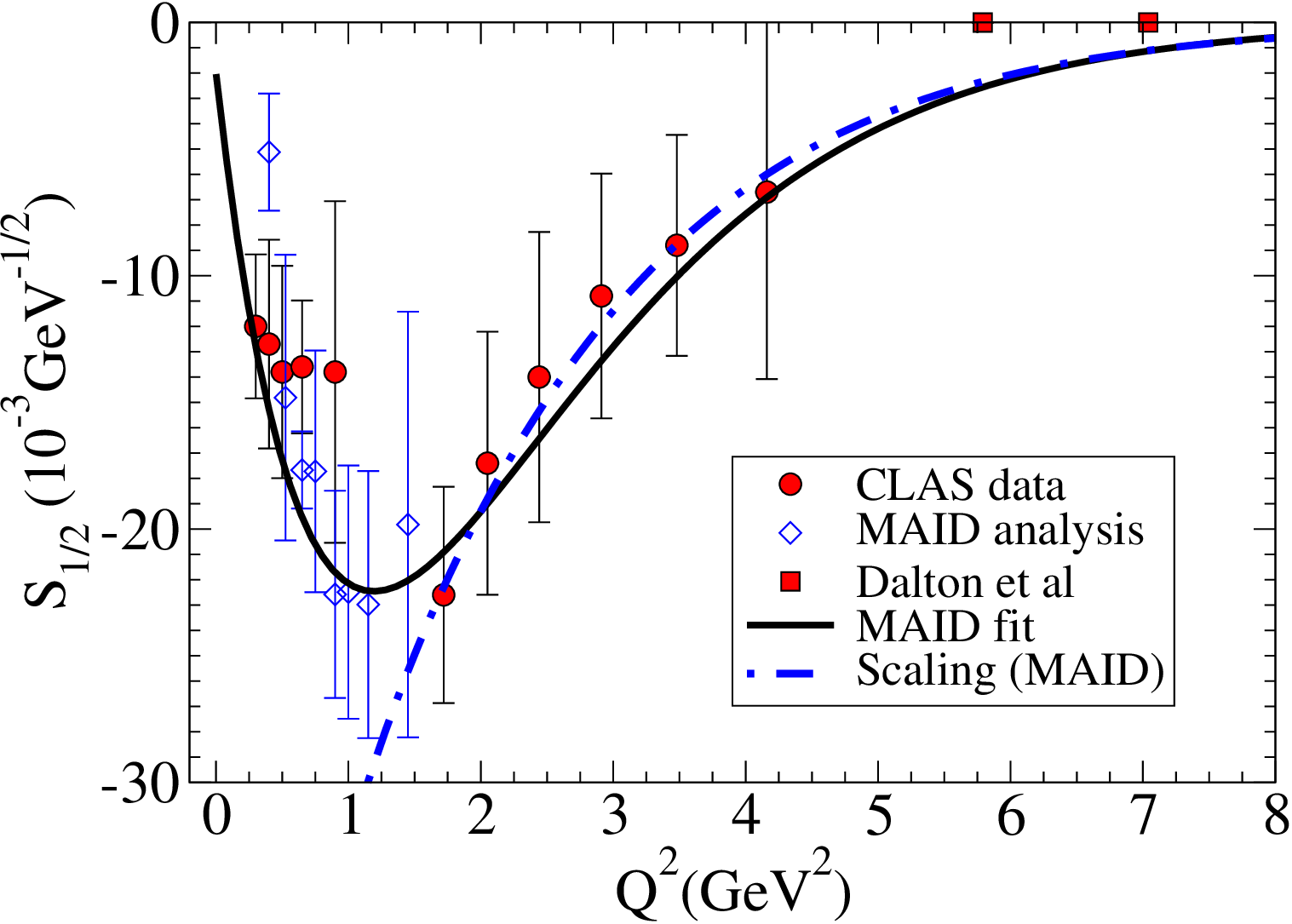} } }
\caption{\footnotesize
$\gamma N \to N(1535)$ helicity amplitudes compared
with the MAID parametrization.
Data are from Refs.~\cite{CLAS,MAID,Dalton09}.
The filled-squares corresponding to the data from
Dalton~\cite{Dalton09} for $S_{1/2}$,
are included to emphasize that
$S_{1/2}=0$ is assumed in the determination
of $A_{1/2}$.}
\label{figMAID1}
\end{figure}

%%%%%%%%%%%%%%%%%%%%%%%%%%%%%%%%%%%%%%%%%%%%%%%%%%%%%%%%%%%%%
% SECTION
%%%%%%%%%%%%%%%%%%%%%%%%%%%%%%%%%%%%%%%%%%%%%%%%%%%%%%%%%%%%%

%\subsection{Scaling using MAID}

{\it Scaling using the MAID fit}:
To test the scaling relation Eq.~(\ref{eqScaling}), we use
the MAID parametrization for the amplitudes
$A_{1/2}$ and $S_{1/2}$.
The MAID parametrization is a fit to the MAID analysis data
that can be extended for the high $Q^2$ region~\cite{MAID}.
In Fig.~\ref{figMAID1} we compare
the CLAS data for $\gamma N \to N(1535)$
with the MAID fit [solid line].
In addition to $S_{1/2}$ we calculate the result
estimated by Eq.~(\ref{eqScaling}) [dash-dotted line].
From the figure we conclude that
the relation~(\ref{eqScaling}) is indeed a good
approximation for MAID parametrization for $S_{1/2}$
in the region $Q^2 > 1.5$ GeV$^2$.
Note that in the figure both the $S_{1/2}$ parametrization and
the results of $S_{1/2}$ derived from the scaling,
are within the errorbars.
One can also see that the
MAID parametrization and the approximation
are indistinguishable for $Q^2 > 5.5$ GeV$^2$.
However, one must be careful in the
extension of the MAID parametrization
for the very high $Q^2$ region, since
the parametrization is based
on the analytical expression regulated
by exponential functions such as $e^{-\beta Q^2}$.
This is successful in the intermediate
$Q^2$ region data, but
differs asymptotically from the expected
power law behavior predicted by pQCD,
and also from the partial scaling suggested
by the data from other reactions.
Finally note the differences in scales for $A_{1/2}$ and $S_{1/2}$
in Fig.~\ref{figMAID1}.

%%%%%%%%%%%%%%%%%%%%%%%%%%%%%%%%%%%%%%%%%%%%%%%%%%%%%%%%%%%%%
% SECTION
%%%%%%%%%%%%%%%%%%%%%%%%%%%%%%%%%%%%%%%%%%%%%%%%%%%%%%%%%%%%%

%\subsection{Spectator quark model}

{\it Spectator quark model}:
We consider now the $\gamma N \to N(1535)$ reaction based
on a constituent quark model.
By this, we intend to demonstrate the usefulness
of the {\it scaling relation}, and
shed some light on the underlying physics.
The use of a quark model
instead of a phenomenological parametrization,
has an advantage to relate the obtained results
with the underlying physics.
In the present case we can decompose the contributions
for the form factors from the valence quark structure
and those from the quark-antiquark excitations,
which are interpreted
as meson cloud excitations in the low $Q^2$ region.
In Ref.~\cite{S11} it was shown that
the spectator quark model
predictions for $F_2^\ast$ are consistent with
the estimates of the EBAC group for the
contributions from the bare core near $Q^2=2$ GeV$^2$ \cite{EBAC},
when the meson cloud is turned off.
Also in Ref.~\cite{Jido08}, where $N(1535)$ was described
as a dynamically generated resonance
and therefore only meson cloud was taken
in consideration, the contributions for $F_2^\ast$
are negative but with a magnitude close to the results
of the spectator quark model~\cite{JidoPC}.
These results suggest that $F_2^\ast \simeq 0$
in the region $Q^2 > 1.5$ GeV$^2$, 
can be interpreted as the cancellation between
the valence quark effects and those of the meson cloud.
Although the meson cloud contributions are
expected to fall faster than
those of the valence quarks,
they can be significant in some
non-leading order form factors
in the intermediate $Q^2$ region.
An example of the dominance of the meson cloud
effects over the valence quark effects
are the electric and
Coulomb transition quadrupole form factors
in the $\gamma N \to \Delta$ reaction
for $Q^2=0-6$ GeV$^2$~\cite{NDeltaD,LatticeD}.

\begin{figure}[t]
%\vspace{.8cm}
\centerline{
\mbox{
\includegraphics[width=3.0in]{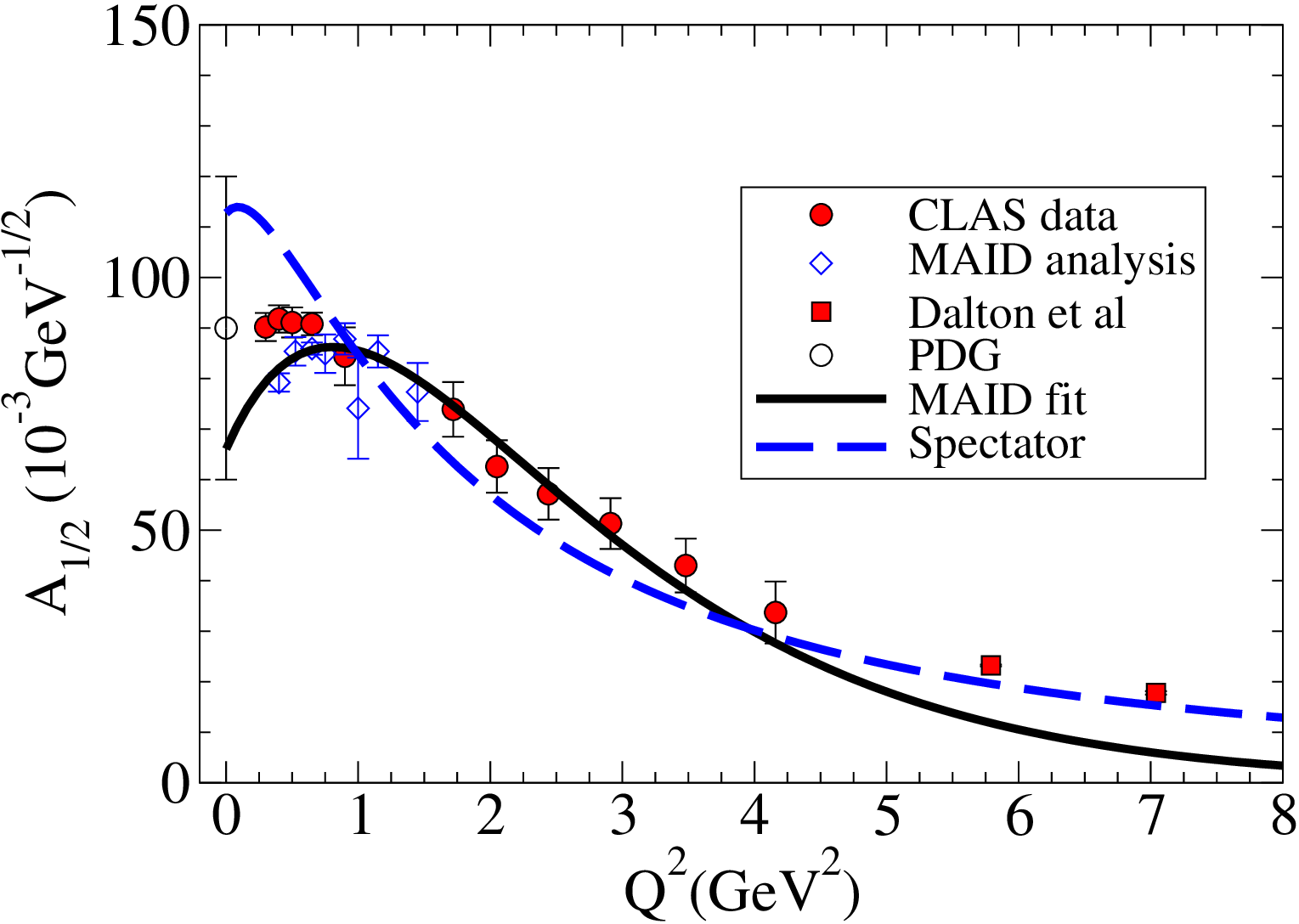} } }
\vspace{.3cm}
\centerline{
\mbox{
\includegraphics[width=3.0in]{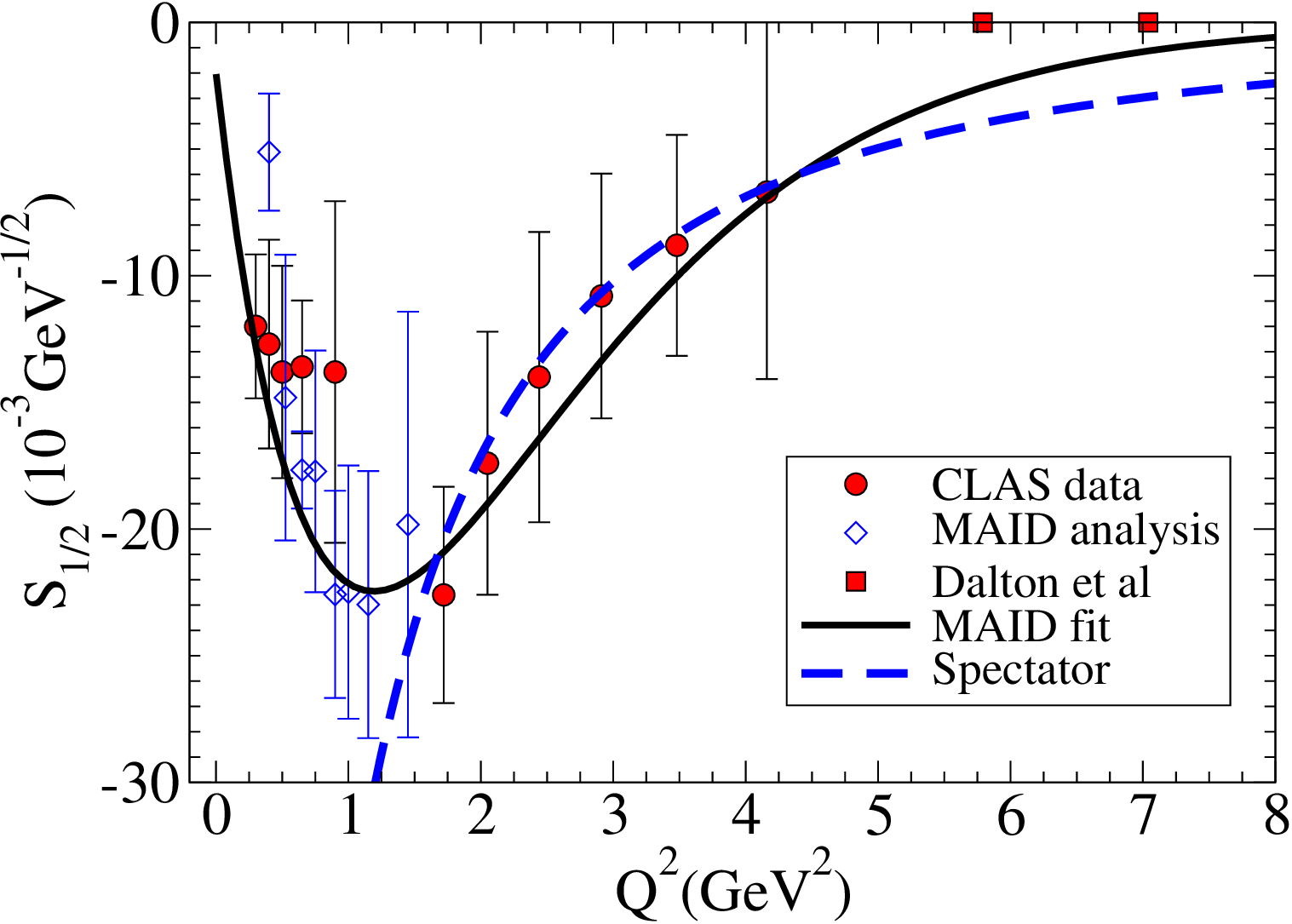} } }
\caption{\footnotesize
$\gamma N \to N(1535)$ helicity amplitudes calculated
in the spectator quark model~\cite{S11} [dashed line].
They are calculated using the results for $F_1^\ast$ and
Eqs.~(\ref{eqSpectator}). Data are from Refs.~\cite{CLAS,MAID,Dalton09}.}
\label{figS11Sp}
\end{figure}

Based on the discussions made above, below
we apply the covariant spectator quark model
developed in Ref.~\cite{S11} to
the $\gamma N \to N(1535)$ reaction in
the region $Q^2 \gg \left(\sfrac{M_S^2-M^2}{2M_S}\right)^2=0.23$ GeV$^2$,
without having any adjustable parameters.
As shown in Fig.~\ref{figS11},
the model [dashed line] gives an excellent
description for the $F_1^\ast$ data
in the region $Q^2 > 2.3 $ GeV$^2$.
This is not surprising, since the valence quark contributions
are in general small in the high $Q^2$ region.
Combining the result for $F_1^\ast$
and the assumption that $F_2^\ast =0$ for $Q^2 > 1.5$ GeV$^2$
(see Fig.~\ref{figS11}), we get:
\ba
& &
A_{1/2}(Q^2)= -2 b F_1^\ast(Q^2), \nonumber \\
%\label{eqSpectator0}\\
& &
S_{1/2}(Q^2)= \sqrt{2} b
\frac{(M_S- M)|{\bf q}|}{Q^2} F_1^\ast(Q^2).
\label{eqSpectator}
\ea
This set of equations is consistent with
the scaling relation Eq.~(\ref{eqScaling})
for $Q^2 > 1.8$ GeV$^2$, but provides also
a method to calculate $A_{1/2}$ through $F_1^\ast$.
The results obtained in the spectator quark model [dashed line]
are presented in Fig.~\ref{figS11Sp}.
One can see the excellent agreement with the helicity
amplitude data in the region $Q^2 > 2.3$ GeV$^2$.
Assuming that the scaling relation, or relations in
Eqs.~(\ref{eqSpectator}),
hold for very high
$Q^2$, we expect that the ratio, $S_{1/2}/A_{1/2}$, converges to
$-\sfrac{1}{\sqrt{2}}\sfrac{M_S-M}{2M_S} \simeq - 0.13$
in the limit $Q^2 \to \infty$.
However, note that the approximation works
only in the region \mbox{$Q^2 \gg (M_S+M)^2= 6.1$ GeV$^2$},
meaning that the convergence is very slow.

%{\it Conclusions}:
{\it Summary}:
In this work we have found
a novel scaling relation
between the $A_{1/2}$ and $S_{1/2}$ helicity amplitudes
for the $\gamma N \to N(1535)$ reaction,
given by Eq.~(\ref{eqScaling}), for $Q^2> 1.8$ GeV$^2$.
The scaling relation is a consequence
of the experimental result for the Pauli-type form factor:
$F_2^\ast \simeq 0$ for $Q^2 > 1.5$ GeV$^2$.
This is very surprising, and has never been observed
in similar reactions like $\gamma N \to N$
or $\gamma N \to N(1440)$. 
The scaling relation between the helicity amplitudes,
found in this work, is also supported by the MAID
parametrization for $Q^2> 1.5$ GeV$^2$.
In a quark model formalism the result
can be interpreted as the cancellation
between the valence quark and meson cloud effects
for $F_2^\ast$.
As a consequence, the helicity amplitudes $A_{1/2}$ and $S_{1/2}$,  
can be simultaneously predicted   
using a valence quark model with the results 
of $F_1^\ast$ for $Q^2 > 1.5$ GeV$^2$. 
We have demonstrated this using the covariant spectator quark
model of Ref.~\cite{S11}, which is valid for $Q^2> 2.3$ GeV$^2$.
We conclude that, although $N(1535)$ may possibly be described as
a dynamically generated resonance~\cite{Kaiser95,Jido08,EBAC},
the transition form factors for $\gamma N \to N(1535)$
can be very well described in
a constituent quark model for high $Q^2$.
We also note that, 
although the scaling relation is consistent with
the spectator quark model and the MAID parametrization,
they give different predictions 
for $Q^2 > 5$ GeV$^2$ (see Fig.~\ref{figS11Sp}).
Then, a precise experimental determination of
the helicity amplitudes,
particularly for $S_{1/2}$ in the high $Q^2$ region,
will be essential to clarify the role
of the helicity amplitudes, and
to test the scaling relation
given by Eq.~(\ref{eqScaling}).

%{\bf Acknowledgments:}

The authors thank Daisuke Jido for sharing the
results from Ref.~\cite{Jido08}.
This work was supported in part by the European Union
(HadronPhysics2 project ``Study of strongly interacting matter'').
This work was also supported by the University of Adelaide and
the Australian Research Council through grant FL0992247 (AWT).
G.~R.~was supported by the Funda\c{c}\~ao para
a Ci\^encia e a Tecnologia under Grant No.~SFRH/BPD/26886/2006.

\end{document}